# Multi-layer scintillation detector for the MOON double beta decay experiment: Scintillation photon responses studied by a prototype detector MOON-1

H. Nakamura[a,b,*], P. Doe[c], H. Ejiri[d,e], S.R. Elliott[c,f], J. Engel[g], M. Finger[h], M. Finger. Jr[h], K. Fushimi[i], V. Gehman[c,f], A. Gorin[j], M. Greenfield[k], V.H. Hai[b], R. Hazama[b,l], K. Higa[b], T. Higashiguchi[b], K. Ichihara[b], Y. Ikegami[b], J. Imoto[b], H. Ishii[b], T. Itahashi[b], H. Kaneko[b], P. Kavitov[m], H. Kawasuso[i], V. Kekelidze[n], K. Matsuoka[b], T. Mizuhashi[b], D. Noda[b], M. Nomachi[b], K. Onishi[b], T. Ogama[b], A. Para[o], R.G.H. Robertson[c], M. Sakamoto[b], T. Sakiuchi[b], Y. Samejima[b], Y. Shichijo[b], T. Shima[d], Y. Shimada[b], G. Shirkov[n], A. Sissakian[n], M. Slunecka[h], Y. Sugaya[b], A. Titov[n], M. Uenoyama[b], S. Umehara[b], A. Urano[b], V. Vatulin[m], V. Voronov[n], J.F. Wilkerson[c], D.I. Will[c], K. Yasuda[i], S. Yoshida[p] and M. Yoshihuku[b]

[a]*NIRS, National Insititue of Radiological Sciences, Chiba, 263-8555, Japan*
[b]*OULNS, Physics, Osaka University, Toyonaka, Osaka 560-0043, Japan,*
[c]*CENPA, University Washington, Seattle, WA 98195, USA,*
[d]*RCNP, Osaka University, Ibaraki, Osaka 567-0047, Japan,*
[e]*JASRI-Spring8, Mikazuki, Hyougo 679-5198, Japan,*
[f]*LANL, P.O.Box 1663, MSH 803, Las Alamos, NM 87545, USA,*
[g]*Physics and Astronomy, Univ. North Carolina, Chapel Hill, NC 27599,USA,*
[h]*Charles University, FMP, CZ-18000 Praha 8, Czech Republic,*
[i]*IAS, University of Tokushima, Tokushima 770-8592, Japan,*
[j]*IHEP, 142284 Protvino, Russia,*
[k]*Physics, International Christian University, Tokyo 181-8585, Japan,*
[l]*Hiroshima University, Higashi Hiroshima, Hiroshima 739-8527, Japan,*
[m]*VNIIEF, Sarov, Russia,*
[n]*Joint Institute for Nuclear Research, 141980 Dubna, Russia,*
[o]*FNAL, P.O.Box 500, Batavia, IL 60510-0500, USA,*
[p]*RCNS, Tohoku University, Sendai 980-8578, Japan*

**Abstract**

An ensemble of multi-layer scintillators is discussed as an option of the high-sensitivity detector Mo Observatory Of Neutrinos (MOON) for spectroscopic measurements of neutrino-less double beta decays. A prototype detector MOON-1, which consists of 6 layer plastic-scintillator plates, was built to study the sensitivity of the MOON-type detector. The scintillation photon collection and the energy resolution, which are key elements for the high-sensitivity experiments, are found to be 1835±30 photo-electrons for 976 keV electrons and $\sigma$ = 2.9±0.1 % ($\Delta E/E$ = 6.8±0.3 % in FWHM) at the $Q_{\beta\beta}$ ~ 3 MeV region, respectively. The multi-layer plastic-scintillator structure with good energy resolution as well as good background suppression of $\beta-\gamma$ rays is crucial for the MOON-type detector to achieve the inverted hierarchy neutrino mass sensitivity.

## 1. Introduction

Double beta decays (DBD) are sensitive and realistic probes to study the Majorana nature of neutrinos (ν), the absolute ν–mass scale and the ν–mass spectrum. Recent neutrino-studies suggest that the effective ν-mass to be studied by neutrino-less double beta decay (0νββ) is of the orders of 100~1 meV, depending on the mass spectrum, if the neutrino is a Majorana particle. Actually, it may be of the order of 100, 30, or 2 meV, if the mass spectrum is with quasi-degenerate (QD), inverted hierarchy (IH), or normal hierarchy (NH). Thus it is of great interest to study 0νββ with the mass sensitivity of the order of 10 meV.

Current detectors for 0νββ experiments are limited by the mass sensitivity of the order of 200 meV in the QD region because of the limited total ββ isotopes to be used for their experiments. High-sensitivity 0νββ experiments with mass sensitivities of the QD mass of 100 meV and the IH mass of 30 meV may require quantities of ββ isotopes on the orders of 0.1-ton and 1-ton, respectively.

Several experiments are planned to study the effective mass in the QD and IH mass regions. Majorana/GERDA, CUORE/CUORICINO, EXO, XMASS CANDLES and others use calorimetric detectors, which are themselves made of ββ isotopes. On the other hand, MOON, NEMO/Super-NEMO and DCBA use spectroscopic detectors, with ββ sources that are positioned within but not an integral part of the detection system. Detailed discussions on ν-masses to be studied by DBD, ν-masses suggested by ν-oscillations and present and future DBD experiments are found in review articles and references therein [1], [2], [3], [4], [5], [6].

The present paper is concerned with the MOON-type spectroscopic detector [7]. Mo Observatory Of Neutrinos (MOON) is based essentially on the ELEGANT V detector [8], but is expanded to improve the mass sensitivity by 2 orders of magnitudes for studying the effective mass in the QD and IH mass regions. Merits of the MOON-type spectroscopic experiment are as follows.

1. Spectroscopic studies of ββ energy and angular correlations enable one to identify the ν-mass term.

2. Since ββ sources are separated from detectors, one can select such ββ isotopes with large $Q_{\beta\beta}$ values and nuclear responses (matrix elements) that give sufficient nuclear sensitivity to enhance the 0νββ signal rate and to place the 0νββ signal well above most BG ones.

3. It is realistic to build a high-sensitivity detector with mass sensitivity of the order of 10 meV using ton-scale ββ isotopes provided that it has energy resolution better than about $\sigma \approx 3\%$ ($\Delta E/E \approx 7\%$ in FWHM) at the $^{100}$Mo $Q_{\beta\beta}$ value (3.034-MeV).

A prototype detector, MOON-1, was built to study the ν-mass sensitivity of the MOON-type detector. In the present report, we discuss mainly the photon responses, i.e. the scintillation photon collection, the energy resolution and the energy calibration, which are key elements for high-sensitivity experiments. Other options of the MOON detector will be discussed elsewhere.

The ν-mass sensitivity is discussed briefly in section 2. The MOON detector configuration and the prototype detector, MOON-1, are described in section 3. The scintillation photon collection, the energy resolution and the energy calibration are discussed in detail in section 4. Concluding remarks are given in section 5.

## 2. Neutrino mass sensitivity and ββ detector

The limiting effective ν-mass $<m>$ which can be studied by 0νββ experiments with $^{100}$Mo over a 5 year period is known to be given by the 0νββ nuclear sensitivity $S_N$ as follows [1].

$$\langle m \rangle = 12/\sqrt{S} \ meV \quad (1)$$

$$S = S_N \times \sqrt{n_{\beta\beta}} / \sqrt{B} \times \varepsilon \quad (2)$$

, where $n_{\beta\beta}$ is the ββ isotopes in unit of ton, $B$ are the BG rates per ton per year of the ββ isotopes at the $^{100}$Mo $Q_{\beta\beta}$ value (3.034-MeV), and $\varepsilon$ is the detection efficiency of the 0νββ event after all kinds of hard and soft cuts. The nuclear sensitivity $S_N$ in unit of $10^{-24}$ y$^{-1}$(eV)$^{-2}$ is expressed as $S_N = G/|M|^2$ with $G$ and $M$ being the phase-space factor and the nuclear matrix element in unit of the electron mass, respectively. The mass sensitivity in equation (1) in ref [1] is based on the 2σ confidence level to describe detector sensitivity, yet a 4-5σ confidence level will be required for 0νββ claims. In case of MOON with the detection efficiency of around $\varepsilon \approx 0.3$ and the nuclear sensitivity $S_N \approx 1$, one gets $<m> \approx 22 \ (B/n_{\beta\beta})^{1/4}$ meV. Then one needs low-background ($B \approx 1$ per ton year) and large-scale detectors with ton-scale ββ isotopes ($n_{\beta\beta} \approx 1$) to achieve the IH sensitivity of $<m> \approx 30$ meV. Since the energies of 0νββ signals for these ββ isotopes exceed the energies of most β–γ signals from the natural and cosmogenic RI impurities, their BG rates may not be sufficiently realistic. Then major BG events are due to the tail of the 2νββ spectrum in the 0νββ window, which depend strongly on the 0νββ energy window and thus on the energy resolution.

Key points of spectroscopic experiments with the mass sensitivity of an order of IH mass (~ 30 meV) are to build a large low-background detector to accommodate ββ isotopes of the order of $n_{\beta\beta} \approx 1$ (ton) and to achieve the sufficient energy resolution with $\sigma \approx 3\%$ ($\Delta E/E \approx 7\%$ in FWHM) at the $^{100}$Mo $Q_{\beta\beta}$ value (3.034-MeV) regions to reduce the 2νββ BG rate to the order of $B \approx 5$ (per ton year). It is really a challenge to meet these conditions in the few-MeV energy region.

## 3. MOON detector concept

MOON consists of multi-layer detector modules [7], [9], [10]. Each module of the plastic-scintillator option MOON is composed of a plastic-scintillator plate, two thin detector planes for position and particle identifications, a thin ββ source film interleaved between the two planes. The ββ vertex point is identified by the detector planes for position and particle identification, while energies of the two β rays are measured by two adjacent plastic-scintillator plates. All other modules (layers) are used as active shields to reject γ rays in the MOON detector. Scintillation photons are collected by photo-multiplier tubes (PMT) around the plastic-scintillator plate. Assuming one module is composed of one plastic-scintillator with ~ 1×1×0.01 m$^3$ and a thin ββ source film with 100×100 cm$^2$ and 20 mg/cm$^2$, the detector has 0.2 kg of ββ isotopes per module. Thus one unit made of 200 modules at ~1cm intervals has 40 kg of ββ isotopes with the total detector volume of the order of ~ 1×1×4 m$^3$ in case of scintillation-fiber planes for particle identification. The MOON-type spectroscopic experiment has several unique features.

1. Individual ββ rays emitted in opposite directions are measured in coincidence by two adjacent plastic-scintillator plates to confirm the ν-mass term in the 0νββ.

2. The multi-module structure makes it realistic to build a compact detector of the order of ~ 0.1 m$^3$ per kg ββ isotopes to accommodate a ton scale of ββ isotopes.

3. Good energy resolution of $\sigma \approx$ 3 % ($\Delta E/E \approx$ 7 % in FWHM) at the $^{100}$Mo $Q_{\beta\beta}$ value (3.034-MeV) can be obtained by efficient scintillation photon collection to reduce the 2νββ contribution down to the order of 5 per ton year.

4. The multi-layer module structure with a good position resolution enables one to select 0νββ signals and reject RI-background signals [1], [7], [9], [10]. Then BG's from RI isotopes in ββ source with realistic ppt (parts per trillion) purity-level may be quite small.

5. Since the source is separated from detector, one can select the best one or two ββ nuclids from view points of the nuclear matrix element, the phase space, the signal energy, and the 2νββ rate.

Accordingly it is realistic to carry out high-sensitivity ββ experiments in the QD-IH mass region. The present paper aims at demonstrating experimentally the key points of 2 and 3 given above using a prototype MOON-1 detector [10], [11], [12], [13], [14].

## 4. A prototype MOON-1 detector

### 4.1. MOON-1 detector configuration

A prototype MOON-1 detector was constructed to study the scintillation photon responses (photon collection and energy resolution), the feasibility of the multi-layer structures of plastic-scintillator plates and ββ-source films, and the BG-rejection capability. These are crucial points for the high-sensitivity experiment, and thus the results of MOON-1 can be used to prove feasibility of MOON with the mass sensitivity of around 30 meV.

The MOON-1 detector consists of 6-layer plastic-scintillator plates, each with 53×53×1 cm$^3$, as shown schematically in Fig. 1 [13], [14]. RP-408 (BC-408 equivalent) plastic-scintillator plates were provided by REXON. The plastic-scintillator plates are realistic detectors from view points of the low RI impurity, the good photon yield of around 10$^4$ per MeV, and the low-cost of a large quantity on the order of 10 tons.

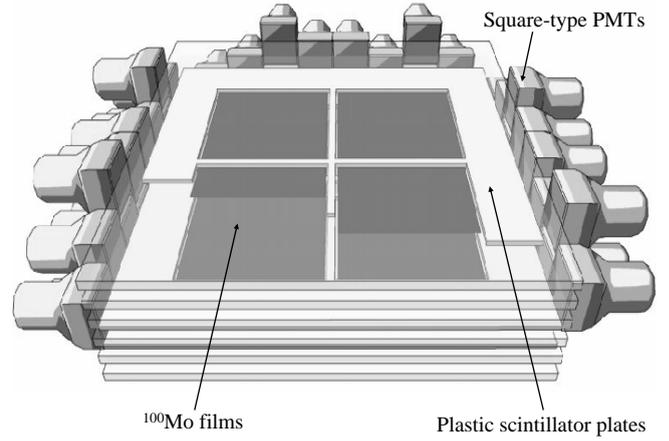

**Fig. 1.** Schematic view of MOON-1 detector with 6-layer plastic-scintillator plates and 3 $^{100}$Mo films. The $^{100}$Mo files are interleaved between two plastic-scintillator plates. The total amount of the $^{100}$Mo films is 142 g. The plastic scintillator plates are viewed by square-type PMT (R6236-01-KMOD provided by Hamamatsu Photonics). The 56 PMTs are coupled to the four sides of the 6 plastic scintillator plates.

The 94.5 % enriched $^{100}$Mo films [15], each having dimensions of 48×48 cm$^2$ ×40 mg/cm$^2$, are interleaved between the plastic-scintillator plates in the MOON-1 detector. The $^{100}$Mo films are covered with aluminized Mylar having surface areas of 53 by 53 cm$^2$ and thicknesses of 6 µg/cm$^2$. The aluminized Mylar suppress the photon cross talk between the adjacent plastic-scintillator plates and support the $^{100}$Mo films.

The 6-layer plastic-scintillator plates are viewed by 6×6 cm$^2$ square-type photo-multiplier tubes (PMT) R6236-01-KMOD provided by Hamamatsu Photonics. The PMT has a K-free window with 0.7-Bq $^{40}$K. The 56 PMTs are coupled to the four sides of the 6-layer plastic-scintillator plates covering about 82 % of the side-surfaces. The silicon cookie, which is made from a silicon rubber SE1821 provided by TORAY, is used as optical connection with ≈ 3 mm thickness. One PMT collects photons from 3 plastic-scintillator plates and the hit plastic-scintillator plate is identified by the PMT hit pattern.

### 4.2. Photon response of plastic-scintillator

Photon response for the same plastic-scintillator as used for the MOON-1 detector was studied using RI sources. The 32 PMTs are coupled to the four sides of plastic-scintillator to get the same coverage as in MOON-1. As the first step, the photon response for each PMT was measured using photons from LED (NSPB500S, 475 nm, NICHIA). The measured spectrum was analyzed in terms of the Poisson distribution for the photo-electron fluctuation and the Gaussian distribution for the PMT gain fluctuation. The PMT response is obtained to be 3.6±0.07 ADC channels per photo-electron. The error is due to statistical fluctuations estimated via peak fitting.

The photon response for each of the plastic-scintillators was measured using the 976-keV K conversion-electron from a $^{207}$Bi source at the center of each plastic-scintillator. The number of photo-electrons for each PMT was deduced from the observed 976-keV peak channel using the measured response of 3.6 channels per photo-electron. The number of photo-electrons depends upon the geometrical position (solid angle) of the PMT with respect to the source, as shown in Fig. 2 (a), (b), (c), and (d). They are plotted for 4 PMT geometries in Fig. 3. The number of photo-electrons are nearly the same among PMTs in approximately the same relative geometrical position ((a), (b), (c), or (d) in Fig. 2). They scatter little around the average value (line in Fig. 3), depending on the photo contact between PMT and plastic-scintillator. Then slight adjustments were made among the 8 PMTs having the same relative geometry.

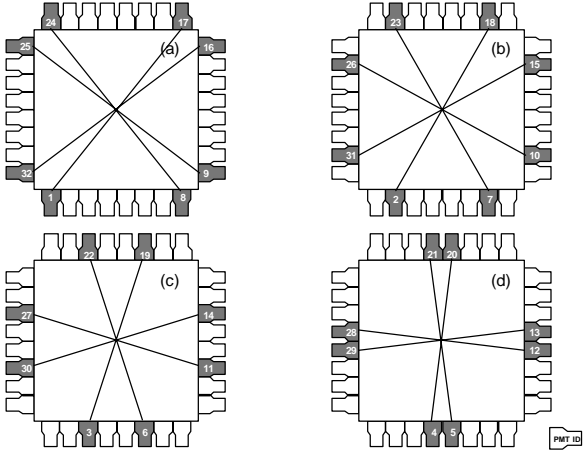

**Fig. 2.** PMT geometry. (a),(b),(c) and (d) shows 4 kinds of the PMT geometrical positions. 8 PMTs in each geometry are symmetrical with respect to the test source at the center. PMT ID numbers are attached on PMT.

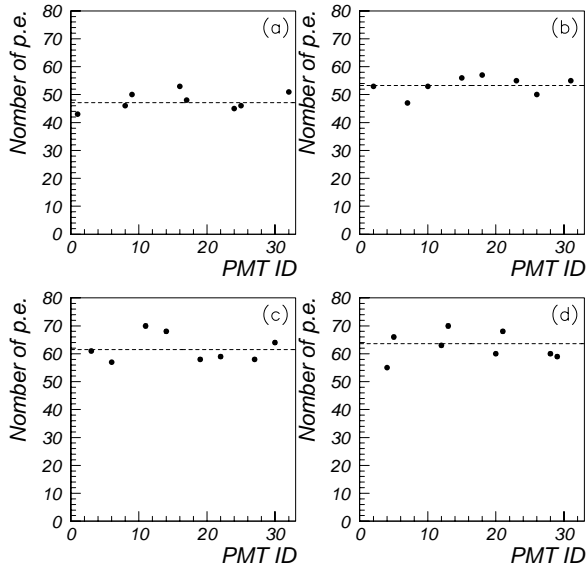

**Fig. 3.** Photo-electron numbers obtained by each PMT. The $^{207}$Bi source with 976-keV K conversion-electrons is set at the center of plastic-scintillator.

Then the energy spectrum is obtained for the K and L conversion-lines from $^{207}$Bi by summing up the energy signals from the 32 PMTs. The obtained spectrum is shown as a function of the number of the photo-electrons in Fig. 4. The number of total photo-electrons is 1830±35 for the 976-keV electron line, which corresponds to 1860 photo-electrons per MeV. This is just what is expected from the total reflection rate for the case of the total number of $10^4$ photons per MeV for the present plastic-scintillator [16] and for the amount of coverage by PMTs and the photo-electron conversion coefficient of 0.25 for the PMTs [17]. The energy resolution of the plastic-scintillator is derived by fitting the observed spectrum in terms of two Gaussian peaks of 976-keV K and 1048-keV L-lines. Here the relative K and L-peak yields are known and the energy resolution is assumed to follow the $\sigma/\sqrt{E}$ dependence. The energy resolution is found to be $\sigma$ = 4.8±0.2 % ($\Delta E/E$ = 11.4±0.5 % in FWHM) at 976-keV, as shown in Fig. 4.

The number of total photo-electrons does not depend within a few % on the source position, as shown in Fig. 5. Since the source position can be derived from relative yields of the photo-electrons at the four sides, one can correct for the slight position dependence.

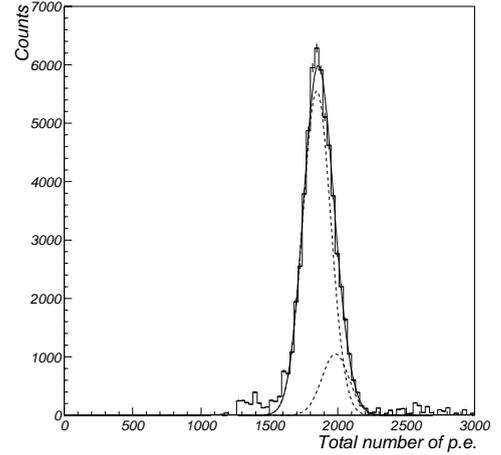

**Fig. 4.** Energy spectrum of the K and L conversion-electrons from $^{207}$Bi by summing up the energy signals from the 32 PMTs with the two-Gaussian peak fit and the individual K and L peak fits.

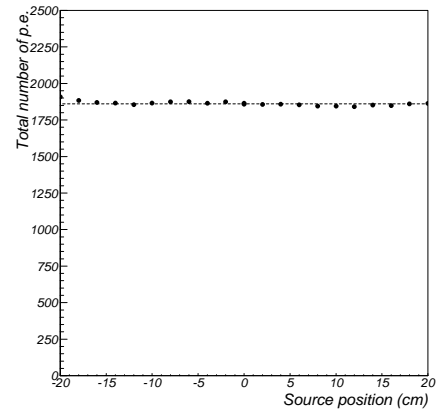

**Fig. 5.** Position dependence of the total number of photo-electrons from the 976-keV $^{207}$Bi K-line. The source position (x axis) gives the distance along the x axis (x, y = 0) from the center (x = 0, y = 0).

### 4.3. MOON-1 detector setting

The MOON-1 detector was set in the active and passive shields of ELEGANT V [8], [18], as shown in Fig. 6. The drift chamber was replaced by the multi-layer plastic-scintillator plates with $^{100}$Mo films. The active shields of 14 NaI(Tl) detectors, each having dimensions of 102×10.2×10.2 cm$^3$, were used to veto γ rays associated with RI backgrounds. The passive shields are 10 cm thick copper bricks (Oxygen Free High Conductivity) and 15 cm thick lead bricks. The MOON-1 and the NaI(Tl) detectors were set in an air-tight box (69.6 cm high and 147.0 cm by

200.0 cm) in order to keep the radon (Rn) concentration low by flushing it with Rn free $N_2$ gas. The experiment was carried out at the Oto underground laboratory with 1400 m water-equivalent depth.

Data taking is made using four kinds of trigger signals from plastic-scintillator, LED, Clock, and NaI(Tl). The first three triggers were taken simultaneously. The plastic-scintillator trigger with the threshold less than 200-keV is used to measure the two β rays by plastic-scintillators. The UV LED trigger is used to monitor the MOON-1 detector. The Clock trigger is used to obtain the pedestal for each PMT. The NaI(Tl) trigger is used to measure γ ray backgrounds. The data are recorded by ORed triggers. The ORed rate is around 11 Hz, with 8 Hz, 2 Hz and 1 Hz from plastic-scintillators, LED, and Clock, respectively. The dead time is 1.2 msec (~ 1 %).

The front end of the data acquisition system is designed using CAMAC at the Oto underground laboratory. The CC/NET (TOYO) [19], which has an internal CPU, is used as the CAMAC crate controller. The Linux OS is installed in the CC/NET. The collected-data program runs on the Linux OS. The data, which are saved in the server computer, are transferred to the computer in RCNP (Research Center for Nuclear Physics) of Osaka University and the quality of which may be checked via the network.

The High Voltage (HV) are supplied from CAEN SY 527 [20] and SY 403 [21] for the PMTs in the MOON-1 plastic-scintillators and the NaI(Tl) detector, respectively. They can be controlled and monitored by the Linux server through RS232C. Therefore, one can access the server from Osaka university.

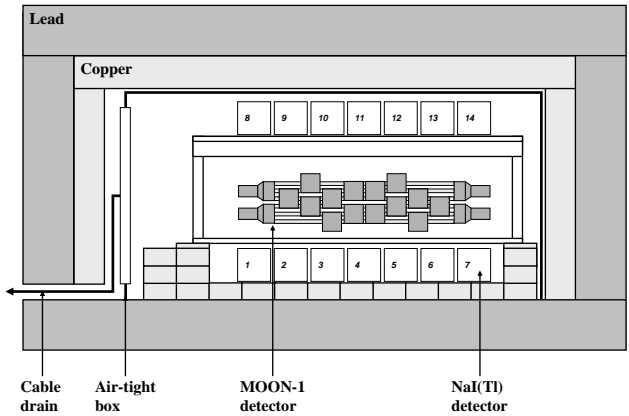

**Fig. 6.** The experimental setup of MOON-1 detector is shown. The MOON-1 detector is set in the active and passive shield. The 14 NaI(Tl) detector is used as the active shield. These detectors are placed in the air tight box to keep the Rn concentration low. The copper and lead bricks are used as the passive shield.

## 5. Results of MOON-1 detector responses

### 5.1. Beta event selection

The present work demonstrates the feasibility of the multi-layer plastic-scintillator modules in terms of having both sufficient scale and energy resolution, which are key elements for such a high-sensitivity experiment. Here, single-layer hit events at the PL3 (the third layer from the top) are selected to estimate the energy resolution of the MOON-1 plastic-scintillator plates.

The experimental setup is shown in Fig. 7. The 1.274-MeV γ rays from a $^{22}$Na source are set at 90 mm above the top of plastic-scintillator. One PMT collects photons from 2 or 3 layers of plastic-scintillators. The selection of the single-layer PL3 event is made by requiring signals from PMTs viewing the PL3 which are not vetoed by signals from other PMTs, as shown schematically in Fig. 8. The energy deposit on PL3 is obtained by summing up signals from PMTs viewing PL3. Here the threshold for the summed signal is set at 200-keV, while that for the veto signal at 50-keV.

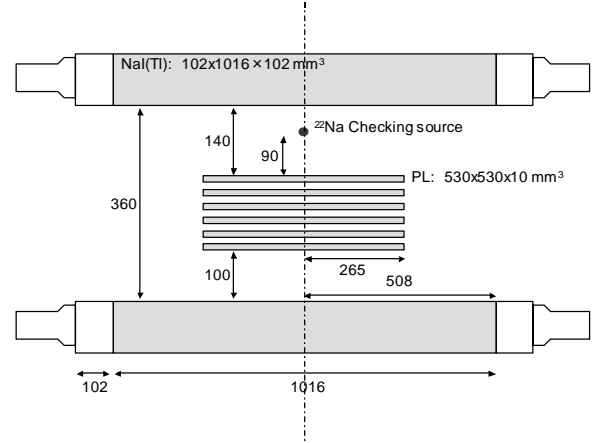

**Fig. 7.** The experiment setup is shown to perform the energy calibration using $^{22}$Na source at 90 mm above the top of the plastic scintillator.

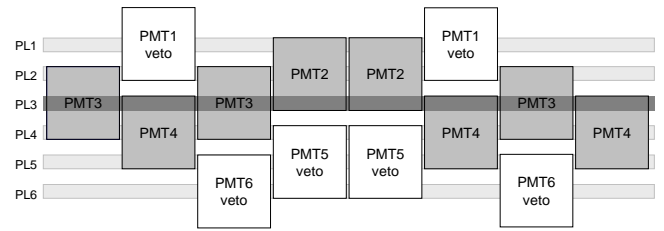

**Fig. 8.** Plastic-scintillator and PMT configuration. Single-layer hit events are selected by requiring signals from all PMT2, PMT3 and PMT4 in coincidence and veto signals from any PMT1, PMT5, or PMT6.

### 5.2. γ ray reconstruction and energy calibration

Compton-scattering is a dominant process for γ rays in plastic-scintillator. Thus the full energy of the gamma peak may be reconstructed by summation of the energy of the Compton-scattered electron deposited in a plastic-scintillator with the corresponding energy of the Compton-scattered γ ray deposited in one of the NaI(Tl). Coincidence between a single-layer hit at PL3 and corresponding energy deposited within NaI(Tl) ID4 is required for each of the 1.274-MeV γ rays from the $^{22}$Na source shown in Fig. 9. The other NaI(Tl) detectors are used as veto counters. Therefore, the two 511-keV γ rays from the source are emitted outside the MOON-1 detector.

The line in the plot of PL3 versus NaI(Tl) ID4 in Fig. 10 is indicative of correlated signals having a total energy of

1.274-MeV. The reconstructed 1.274-MeV γ ray spectrum is obtained by summing the two signals from PL3 and NaI(Tl) ID4, as shown in Fig. 11.

Here, the relative and absolute energy calibrations are obtained via two distinct procedures. One is the relative calibration. The other one is the absolute energy calibration. The relative gains for PMTs are calibrated using the Compton edge of 1.274-MeV γ rays from the $^{22}$Na source. Their signals obey the measured distribution as obtained in the section 4.2. The energy calibration of NaI(Tl) ID4 is performed using the γ rays from the source ($^{22}$Na 511-keV, 1.274-MeV) and the radioactive isotopes ($^{40}$K 1.461-MeV, $^{208}$Tl 2.615-MeV). The reconstructed total energy peaks are used for the absolute energy calibrations of the plastic-scintillators. This method is used for other plastic-scintillator layers as well.

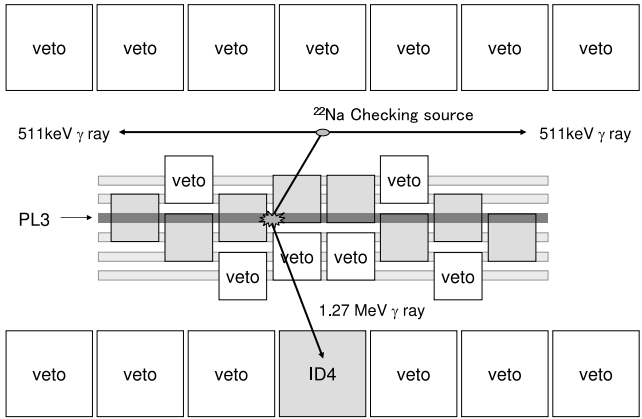

**Fig. 9.** Compton-scattering of $^{22}$Na γ rays. The γ ray source is set on the rough center of a plastic scintillator. The single-layer hit event of the plastic scintillator (PL3) with NaI(Tl) detector (ID4) is selected for the 1.274-MeV γ rays from $^{22}$Na source.

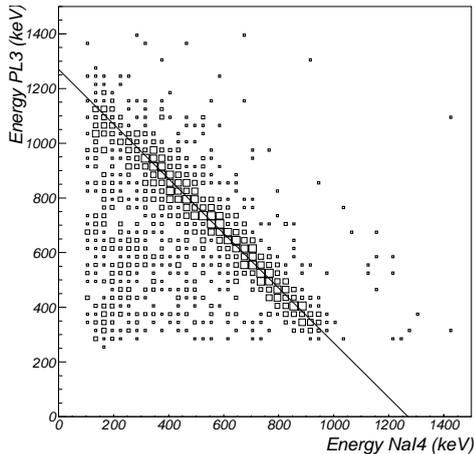

**Fig. 10.** The energy deposit on the plastic scintillator (PL3) and the NaI(Tl) ID4 detector is shown for the event of the 1.274-MeV γ rays from $^{22}$Na source. The line shows that the sum energy of 1.274-MeV.

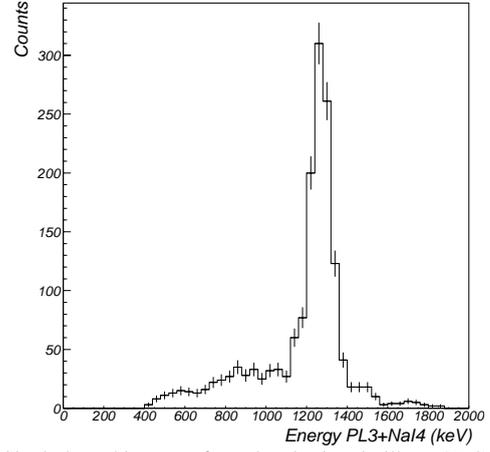

**Fig. 11.** Single-layer hit events from the plastic scintillator (PL3). The full energy peak of this γ ray is made by summing the energy deposits on the plastic scintillator (PL3) and the NaI(Tl) ID4 detector.

### 5.3. Energy resolution of plastic-scintillator

The reconstructed γ peaks from the plastic-scintillator and the NaI(Tl) signals are used to evaluate the energy resolution of the MOON-1 plastic-scintillator plates. Here, the energy and the energy resolution for the reconstructed γ ray are expressed as

$$E_\Sigma = E_{PL} + E_{NaI} \quad (3)$$
$$\Delta E_\Sigma^2 = \Delta E_{PL}^2 + \Delta E_{NaI}^2 \quad (4)$$

where $E_\Sigma$, $E_{PL}$ and $E_{NaI}$ are the energy of the reconstructed γ ray, the energy deposit on the plastic-scintillator and the energy deposit on the NaI(Tl), respectively, and $\Delta E_\Sigma$, $\Delta E_{PL}$, and $\Delta E_{NaI}$ are their respective fluctuations.

First, the energy resolutions of the NaI(Tl) detector were found to be $\sigma = 4.5\pm0.2$ % ($\Delta E/E = 10.5\pm0.4$ % in FWHM), $2.8\pm0.1$ % ($6.6\pm0.2$ % in FWHM) for the γ rays from the source ($^{22}$Na 511-keV, 1.274-MeV) and to be $\sigma = 2.7\pm0.1$ % ($\Delta E/E = 6.4\pm0.2$ % in FWHM), $2.0\pm0.1$ % ($4.6\pm0.3$ % in FWHM) for the radioactive isotopes ($^{40}$K 1.461-MeV, $^{208}$Tl 2.615-MeV) respectively.

The energy window of NaI(Tl) is selected in order to estimate the energy resolution of plastic-scintillator. Requiring a coincidence between the 511-keV energy deposited on the NaI(Tl) ID4 with the 763-keV energy deposit from the Compton-scattered electron on the PL3, the energy resolution of the PL3 at the 763-keV may thus be obtained. The energy window is shown for 1.274-MeV γ rays in Fig. 12.

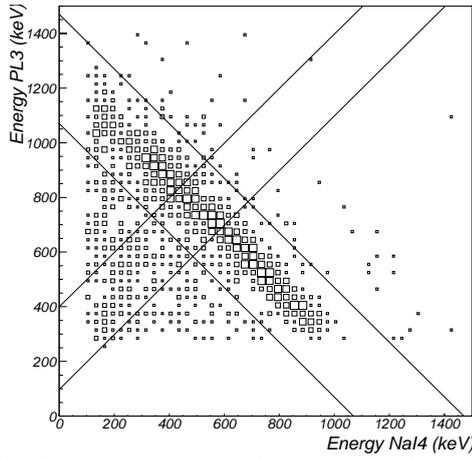

**Fig. 12.** Correlation between energy deposits on the plastic scintillator (PL3) and NaI(Tl) detector (ID4) for the single-layer hit events of the γ rays from $^{22}$Na. The energy window of NaI(Tl) detector is selected at the 511 keV region to estimate the energy resolution for a plastic scintillator at the 1274-511 = 763 keV region.

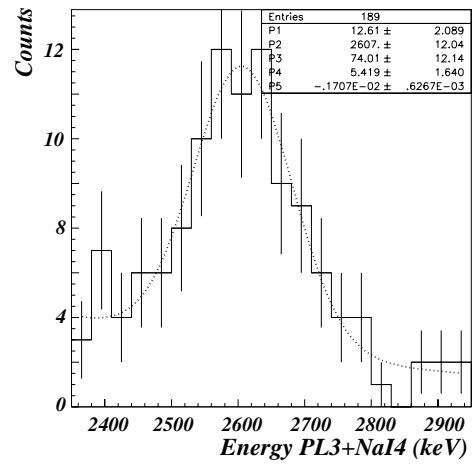

**Fig. 14.** Reconstructed 2.615-MeV γ peaks from $^{208}$Tl at the NaI(Tl) 511-keV energy window by summing the energy of the plastic scintillator (PL3) and the NaI(Tl) detector (ID4). The dotted line is a fit with the energy resolution is $\sigma_{PL+NaI} = 2.9\pm0.1$ % ($\Delta E/E = 6.8\pm0.3$ % in FWHM) at 2.615-MeV region.

The γ ray peaks from the source ($^{22}$Na 1.274-MeV) and the radioactive isotopes ($^{208}$Tl 2.615-MeV) are shown at the energy window $E_{NaI}$ = 511-keV of the NaI(Tl) detector in Fig. 13 and Fig. 14. They are fitted by a Gaussian peak with an exponential background tail. The measured energy deposit at PL3 agrees with the expected value of $E_{PL} = E_{\Sigma} - E_{NaI}$, as shown in Fig. 15. Here the energy scale of plastic-scintillator is calibrated using the 1.461-MeV γ rays from the $^{40}$K radioactive isotope.

The energy resolution for PL3 is found to be well reproduced by $\sigma/\sqrt{E}$ with $\sigma$ = 5.0±0.2 % ($\Delta E/E$ = 11.9±0.5 % in FWHM) in the energy region from 0.5-MeV to 2.1-MeV, as shown in Fig. 16. This energy region covers β ray energies in most spectroscopic ββ experiments. The energy resolutions obtained for the conversion-electrons from the source ($^{137}$Cs 624-keV, $^{207}$Bi 976-keV) agree well with those derived from the reconstructed γ peaks, as shown in Fig. 16. The energy resolution at the $^{100}$Mo $Q_{\beta\beta}$ value (3.034-MeV) is $\sigma$ = 2.9±0.1 % ($\Delta E/E$ = 6.8±0.3 % in FWHM), which is the energy resolution required for MOON with the IH mass sensitivity.

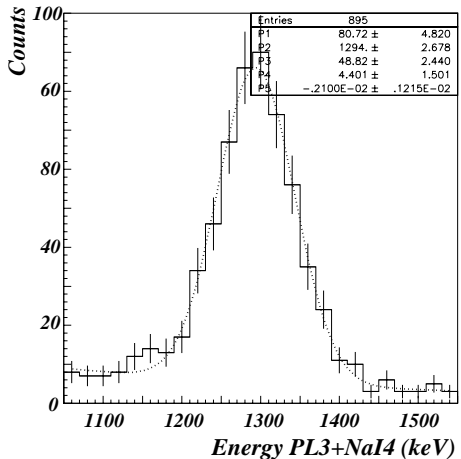

**Fig. 13.** Reconstructed peak of 1.274-MeV γ rays from $^{22}$Na source at the NaI(Tl) 511-keV energy window by summing the energies of the plastic scintillator (PL3) and the NaI(Tl) detector (ID4). The dotted line is a fit with the energy resolution of $\sigma_{PL+NaI} = 3.8\pm0.1$ % ($\Delta E/E = 8.9\pm0.4$ % in FWHM) at 1.274-MeV region.

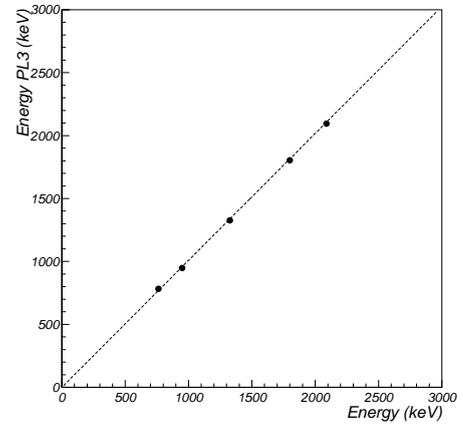

**Fig. 15.** Energy deposit $E_{PL3}$ obtained from the reconstructed γ rays on the plastic scintillator PL3. The statistical error bar is within the data point.

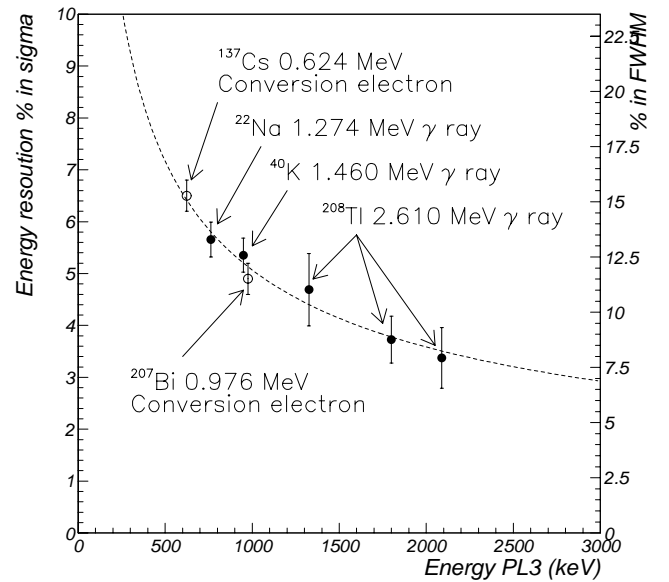

**Fig. 16.** Energy resolution $\sigma$ and $\Delta E/E$ in FWHM of PL3 obtained from reconstructed γ rays. The abscissa axis shows the energy deposit in PL3.

## 6. Concluding remarks

1. MOON is a spectroscopic ββ experiment with a ν-mass sensitivity in the IH mass region (30 meV). It has a multi-layer structure of plastic scintillator plates to accommodate a ton-scale ββ source, a good energy resolution of $\sigma \approx 3$ % ($\Delta E/E \approx 7$% in FWHM) at the 0νββ peak energy, and a good BG rejection capability. These are crucial for achieving the IH mass sensitivity.

2. A prototype detector MOON-1 was built to demonstrate the feasibility and the sensitivity of MOON. It consists of 6 layers of plastic-scintillator plates with dimensions of $53 \times 53 \times 1$ cm$^3$. Photons are collected by 56 PMTs positioned along the 4 sides of the plate covering about 82 % of the plastic-scintillator side-face.

3. The number of total photo-electrons is 1830±35 for the 976-keV electron line. This is just what is expected from the total reflection rate for the case of the total number of $10^4$ photons per MeV for the present plastic-scintillator, the amount of coverage by PMTs, and the photo-electron conversion coefficient of 0.25 for the PMTs.

4. The energy resolution, which is one of the key elements for the large-scale spectroscopic ββ experiments, was measured using conversion-electrons from RI sources as well as Compton-scattering electrons from RI γ rays. A new method of checking the plastic-scintillator response for low-energy electrons is carried out using reconstructed γ peaks via summing the energy of the Compton-scattered electrons in the plastic-scintillators with the energy from the coincident Compton-scattered γ rays in the NaI(Tl).

5. The measured energy resolution is found to be well reproduced by $\sigma/\sqrt{E}$ with $\sigma$ = 5.0±0.2 % ($\Delta E/E$ = 11.9±0.5 % in FWHM) in the energy region from 0.5-MeV to 2.1-MeV. This leads to the energy resolution of $\sigma$ = 2.9±0.1 % ($\Delta E/E$ = 6.8±0.3 % in FWHM) at the $^{100}$Mo $Q_{\beta\beta}$ value (3.034-MeV). This is just what is required for the IH mass sensitivity of around 30 meV.

6. The energy resolution of $\sigma$ = 5.0 % ($\Delta E/E$ = 11.9 % in FWHM) for 1-MeV is worse than $\sigma$ = 2.3 % ($\Delta E/E$ = 5.4 % in FWHM) due to the statistical fluctuation of photo-electrons, since the observed resolution includes the intrinsic (non-statistical) component [22] and others. Proper choice of the plastic-scintillator material and production process may improve the resolution quite a bit.

7. The MOON-1 with 53 cm × 53 cm plastic-scintillator plates works well and can be expanded by a factor 2 to the MOON scale ($\approx$ 100 cm × 100 cm) since the photon attenuation in the plastic-scintillator plate is less than a few percent.

In short, the present work demonstrates the feasibility of the multi-layer plastic-scintillator modules for a high-sensitivity MOON detector in terms of having both sufficient scale and energy resolution, which are key elements for such a high-sensitivity experiment. Another important feature of the detector is the capability of RI background rejection, which is studied experimentally using the MOON-1. MOON is a spectroscopic detector, where ββ source is within but not an integral part of the detector. Thus the MOON-type detector with the multi-layer structure of plastic-scintillator plates can be universally used for ββ decays of other isotopes as well as for other low-energy rare decays.

Experimental studies of the RI-background rejections, the position-sensitive detector plane and others relevant to MOON detectors are reported elsewhere.

## Acknowledgements

The authors are grateful to Osaka University, JASRI, and MSCS for partial support for the present work. The authors are deeply indebted to all other members of the Nomachi group at OULNS. The present work was partially supported by Ministry of Education, Science, Sports and Culture, Grant-in-Aid for Scientific Research (A) 15204022, 2003-2006.

## References

[1] H.Ejiri, 2005, *J. Phys. Soc. Jap.* **74** 2101;
[2] H.Ejiri, 2000, *Phys. Rept.* **338** 265;
[3] J.Vergados, 2002, *Phys. Rept.* **361** 1;
[4] S.Elliott and P. Vogel, 2002, *Ann. Rev. Nucl. Part. Sci.* **52** 115;
[5] J.D.Suhonen and O.Civitarese, 1998, *Phys. Rept.* **300** 123;
[6] F.Simkovic and A.Fassler, 2002, *Prog. Part. Nucl. Phys.* **48** 201;
[7] H.Ejiri et al., 2000, *Phys. Rev. Lett.* **85** 2917;
[8] H.Ejiri et al., 1991, *Phys. lett.* B **258** 17;
[9] P.Doe et al., 2003, *Nucl. Phys.* A **721** 517;
[10] H.Ejiri et al., 2004, *Czech. J. Phys.* **54** B317;
[11] M.Nomachi et al., 2005, *Nucl. Phys. Proc. Suppl.* **138** 221;
[12] H.Ejiri, 2006, *Prog. Part. Nucl. Phys.* **57** 153;
[13] H.Nakamura et al., 2006, *J. Phys:Conf. Ser.* **39** 350
[14] H.Nakamura, 2006, *Dr.thesis, Osaka Univ*;
[15] N.Kudomi et al., 1992, *Nucl. Instrum. Meth.* A **362** 53;
[16] RP408 specifications, REXON Components inc, http://www.rexon.com/RP408.htm;
[17] R6236-01 specifications, Hamamatsu Corporation, http://sales.hamamatsu.com/assets/pdf/parts_R/R6236-01.pdf;
[18] H.Ejiri et al., 1991, *Nucl. Instrum. Meth.* A **302** 304;
[19] CC/NET specifications, TOYO Corporation, http://www.toyo.co.jp/daq/ccnet/
[20] SY527 specifications, Costruzioni Apparecchiature Elettroniche Nucleari spa, http://www.caen.it/nuclear/syproduct.php?mod=SY527;
[21] SY403 specifications, Costruzioni Apparecchiature Elettroniche Nucleari spa, http://www.caen.it/nuclear/syproduct.php?mod=SY403;
[22] H.Murayama et al., 1979, *Nucl. Instrum. Meth.* **164** 447;